Could availability of free articles reduce faculty's dependence on the library? Analysis of items cited by faculty from Singapore Management University

**Nursyeha Binte Yahaya <nursyehay@smu.edu.sg>, Tay Chee Hsien Aaron <aarontay@smu.edu.sg>**

**SMU Libraries, Singapore Management University, Singapore 178901.**

**Purpose:** Recent work suggests that we are close to the 50% mark for freely available articles (Archambault et al., 2013, Jamali and Nabavi, 2015). Given that over 80% of faculty surveyed said that they would search for a freely available version if they do not have immediate access via the library (Wolff et al., 2016) it is natural to wonder if the rise of articles could have reduced our faculty's dependence on the library for access of articles.

**Design:** We sample citations made by researchers who published in 2015 (based on records in the Singapore Management University Institution repository), checked the number of cited papers that were free at the time of the study and then attempted to "carbon date" the freely available papers found at the time of study to determine when they were first actually made available. This allows us to estimate the length of time the cited article was made freely available before the citing paper was published. Knowing this allows us to estimate if the citer could have in theory used the free article for citing purposes.

**Findings:** We find that in our sample of cited papers in Economics, the median freely available cited paper (oldest variant) was made available 7-8 years before the citing paper was published. Based on typical publication lags found in prior studies, 74.3% of the cited papers could be used.

**Research limitations/implications:** The study focused only on faculty from one University in the field of Economics. Also a random sampling method was used to sample citations made and it resulted in a large number of cites to older articles. While this reflects the citation behavior of faculty in the field of Economics(Waller, 2006) and answers our research question on how much they cite could be used, it also means a large number of papers are likely to be free as it is past the embargo. Future studies might want to restrict sampling to newer papers only.

**Originality/value:** This study combines traditional citation analysis with a novel method of "carbon dating" papers and addresses a question of concern to libraries particularly as open access takes hold.

# Introduction

The idea that citation analysis can be used as a journal evaluation and collection development tool has been recognized as far back as in the 70s (Garfield, 1972), and librarians have long done user citation studies for collection development and assessment purposes (Smith, 1981).

For example, Hoffmann and Doucette (2012) reviewed 34 recent studies that use citation analysis methods to inform collection development. Of those 34 studies, researchers have generally chosen to study among other factors (i) The type of item cited, (ii) age of the cited resources and (iii) whether the item cited was in the collection.

While such studies are useful for assessing a libraries' collection, they neglect to take into account the recent rise of open access and free items on the net.

Recent studies that estimate the amount of free articles available online have found this amount to be quite substantial. Estimates range from 20%(Björk et al., 2010) to as high as 61%(Jamali and Nabavi, 2015). Coupled with the fact that document delivery use is down around the world (Boukacem-Zeghmouri et al. 2006; Schöpfel 2015), and that an overwhelming number of researchers (over 80%) will search for free copies when they don't have access (Housewright et al., 2013), there is reason to believe researchers might be using and citing copies of freely available item they find online(Wolff et al., 2016). As such, it may be critical to take into account the availability of free articles.

User citation studies generally have a drawback in that they can only measure availability at the point of study and do not measure retrospective availability.

However unlike library collections which are pretty stable, sources of freely available articles can be very volatile and restrictions such as publisher embargos make it critical to ascertain when the freely available article was first made free to ensure the free article was a theoretically viable source for the citing author to access and use when he was writing the paper.

Hence, this paper seeks to answer the following questions. Firstly, what percentage of items cited by researchers in papers published in 2015 are free to access in 2016 (at the time of the study)? Secondly, how long ago were they made freely available? Lastly, would they be available to be used (at least theoretically) when the citing author was writing his paper?

# Literature Review

Open Access has a history that goes back over 20 years(Suber, 2006) and while there have been disagreements over definitions, Peter Suber defines Open access as such. "Open access (OA) literature is digital, online, free of charge, and free of most copyright and licensing restrictions."(Suber, 2006).

There are generally considered two roads or delivery modes to Open Access. The Gold road which involves providing open access via Journals (typically but not always requiring author processing charges) and the Green Road which provides open access via self-archiving in repositories(Harnad, 2010).

While the best way to achieve open access is a matter of much debate among the scholarly community, from the point of view of a researcher who wants to access the article to read and cite whether an article is available through journals or through a repository is of little interest to them. What researchers need are *articles that are free to access when they need to cite them*.

The literature on open access has grown tremulously particular in the areas studying citation advantage of open access(Eysenbach, 2006) but here we will focus on the strands of research that is perhaps most relevant to our study.

One area of great interest in the open access literature has been tracking and estimating the growth of open access literature.

| Paper | Sample | Coverage of articles checked & time of search | Searched in | Free full text found/estimated | Comment |
|---|---|---|---|---|---|
| Björk et al. (2010) | Drawn from Scopus | 2008 articles searched in Oct 2009 | Google | 20.4% | |
| Gargouri et al. (2012b) | Drawn from Web of Science | 1998-2006 articles searched in 2009. 2005-2010 articles searched in 2011. | "software robot that trawled the web" | 23.8% | |
| Archambault et al. (2013) | Drawn from Scopus | 2004-2011 articles searched in April 2013 | Google and Google scholar | 44% (for 2011 articles) | "Ground truth' of 500 hand checked sample of articles published in 2008, 48% was freely available as at Dec 2012 |
| Martín-Martín et al. (2014) | 64 queries in Google Scholar, collect 1st 1,000 results | 1950-2013 articles searched in May 2014 & June 2014 | Google Scholar | 40% of results | |

| Khabsa and Giles (2014) | Randomly sampled 100 documents from Microsoft Academic Search belonging to each field to check for free version in Microsoft Academic search and Google Scholar | NA , searched in Jan 2013 | Microsoft Academic Search and Google Scholar | 24% (estimated free articles in Google Scholar using capture-release technique) | |
|---|---|---|---|---|---|
| Jamali and Nabavi (2015) | Do 3 queries each in Google Scholar for each Scopus third level subcategory. Check the top 10 results for free full text | 2004–2014 articles, searched in April 2014 | Google Scholar | 61% | |

*Table 1 : Past studies quantifying amount of freely available material on the web.*

Some studies on the growth of open access focus solely on Gold Open Access(Laakso and Björk, 2012), though many now focus on both green and gold access. (Björk et al., 2010, Archambault et al., 2013, Gargouri et al., 2012b). See Table 1 for a summary of some of the later papers.

While the studies focus mostly on sampling from Scopus or Web of Science to check for free availability. Khabsa and Giles (2014) provide a novel strategy to estimate the size of Google Scholar using capture recapture method together with the known size of Microsoft Academic Search. They estimate that Google Scholar has over 100 million documents and 24% of articles on the public web are free.

With regards to Google Scholar, it has become extremely popular among researchers and has been generally recognized as the biggest citation database of scholarly material(Khabsa and Giles, 2014, Orduna-Malea et al., 2015).

This has led to studies that try to estimate the amount of free full text found in Google Scholar (Scott and Sandra, 2014, Jamali and Nabavi, 2015, Martín-Martín et al., 2014). Arguably this can be also seen as an estimate of the amount of freely available material given that the index of Google Scholar is probably the biggest single source of articles.

For example, Martín-Martín et al. (2014), created 64 queries and checked the number of journal articles that were freely available in the top 1,000 results. They found over 40% of items freely available. Jamali and Nabavi (2015) with a much smaller sample found 61% freely available.

While all these studies are useful in establishing the amount of open access or freely available items at the time of the study, they do not directly measure the use or value of such material as they do not take into account citing patterns.

User studies involving citation analysis avoid this issue and show the true value of freely available items. For example, Harder et al. (2015) studied the percentage of free items cited by Wikipedia articles and found that 12.8% of citations to journal articles (from 5,000 English Language Wikipedia) are freely available. Unfortunately this study only considers an item as free if it is available in a limited number of sources such as arXiv and Pubmed central, obviously this vastly undercounts the actual number of cites made to freely available items.

Burns (2013) sampled 999 references from citeulike.org and calculated the possibility that a user could start from Google Scholar and access full text without using a proxy.

While both studies take into account whether an item is cited (or in the former case whether they are collected in citeulike.org) hence are likely to show actual use, they like all studies cited here only consider whether the item is free at the time of the study. As such they cannot answer the question of whether the freely available item was used or usable by the citing author when he wrote his paper.

## Methodology

Our method starts by borrowing from traditional user citation studies by sampling cited papers from our institutional repository INK (http://ink.library.smu.edu.sg/). By sampling from our institutional repository rather than traditional sources like Web of Science, hopefully we can provide a more complete picture of what our faculty are citing. As a first cut, we choose to focus only on citations from journal articles in the disciplines of Economics and Social Science to other journal articles.

While traditionally the next step would be to study how many of these citations are in the collection, in this study we focus on freely available material and ignore the availability of library holdings. As such, we check for freely available versions of the cited items by searching the title of the item in Google Scholar.

While some studies have used Google, Google Scholar as well as custom made bots to establish whether a selected item was freely available (Archambault et al., 2013, Gargouri et al., 2012a), we choose here to use Google Scholar only. We believe this is a fair simulation of the average researcher as most researchers claim to use Google Scholar for finding literature (Nielsen, June 2014) . Of course it is likely not all of this free content discoverable on Google Scholar is technically legal, however it is unlikely the average researcher will notice or worry about it as long as the article he needs is accessible.

Using Google Scholar, we enter the title of the sampled cited papers into the search and check for free versions. Google Scholar typically groups all variants of what it considers the same item together under one entry. We then record the URLs of all variants that are freely available.

**The timing issue**

A critical issue to consider is when the paper was made free.

There is evidence to believe that publishers are starting to lengthen embargo periods which may affect how soon articles are made free via Green Open access.

For example, a study of the original 107 publishers listed on the SHERPA/RoMEO Publisher Policy Database over 12 years found that while the number of RoMEO Green publishers increased slightly, there was a growth in the length of imposed embargo (Gadd and Troll Covey, 2016). This together with the increased volume of restrictions on when self-archiving may take place, it is reasonable to wonder if more and more papers are self-archived later, which might be too late for use by authors who might want to read and cite them.

But how do we determine when an article was made freely available?

While we can easily determine if a cited item is freely available now, it can be tricky to determine if the item was free at the time it was cited.

Articles that are in Gold Open access can probably be safely be considered freely available since inception.

However determining the actual dates when articles made available via Green Open Access was released is much trickier.

Such articles can be found in various avenues. Firstly, one can find articles self-archived by researchers in subject or institutional repositories. Also sites like ResearchGate, academia.edu are becoming top sources of freely available material (Jamali and Nabavi, 2015, Martín-Martín et al., 2014). Lastly lots of free articles resides on personal home pages, commercial webpages etc. In some of these cases (e.g. some institutional repository systems), it might be possible to check the date the item was made available but for most cases it is not easy.

To determine if the author could have used the free version of the paper we found today when he was writing his paper, we would have to determine when the free paper was put up. In other words we have to carbon date the page.

Take in Figure 1

In the above hypothetical example (Figure 1), the cited paper E (published in 2001) was made freely available in 2014, 1 year before the paper that cited it – Citing paper A was published in 2015.

Take in Figure 2

In the next hypothetical example (Figure 2), the cited paper F (published in 2001) was made freely available in 2002, 13 years before the paper that cited it – Citing paper B was published. As such, we define "free citing window" as follows.

> **Free citing window** = Year in which citing paper was published - Year in which cited article was made freely available

In general the bigger the free citing window, the longer the paper has been available for free before the citing paper was published and hence the more likely the author of the citing paper could have used it[1].

Following traditional user citations studies, we also define Citation age as follows:

> **Citation age** = Year in which citing article was published - Year in which cited paper was published

**Carbon dating**

We rely on the work of SalahEldeen and Nelson (2013) to determine when an article was made free. Their work uses a host of ways to try to determine the age of links. These include

a) Bitly for the first time a link was shortened
b) Topsy for the first time a link was tweeted
c) Public web archives (including wayback machine) for the first time a link appears
d) Google's date of last crawl of the page
e) Last-Modified HTTP response header of the resource
f) Backlinks from Google

---

[1] The free citing windows is typically positive but it can be negative if the free version of the cited paper is available freely after the citing paper is published.
For example, imagine a paper A published in 2015 cites paper B. We search Google Scholar for Paper B in June 2016 and find a free copy. This paper is then found to have been made free in Jan 2016. In this case, we have a negative free citing window.

Based on all these methods, they created a model to predict an estimated date.

The method isn't perfect and in their testing against 1,200 "gold standard" links, 24.9% failed to retrieve any date and they had an exact correct prediction (to the day) for 32.78% of links checked. Still this is the only viable way to check, and we use their method as an estimate of how long the URL was up, relying on their "Estimated Creation Date".

**How we carbon dated**

We used the carbon dating service at http://cd.cs.odu.edu/ to carbon date every variant of free article found in Google Scholar to find when they were made free. Due to the inherent uncertainty of the method, we only record the year of the carbon date.

As Google Scholar typically finds more than 1 free copy of the paper, we need to carbon date each variant of the free paper. In the interest of time, we stop carbon dating once we carbon date a variant that is older than 5 years.

At the time of our study only methods a, c), d) and e) was available via the carbon dating service (Topsy had been acquired and was shutdown).

In our sample, we found that of variants that could be carbon dated, 41.1% of the carbon dates came from web archives, 8.9% came from Google's date of last crawl and 51% came from last modified (e)

We use the estimated date prediction (also known as "Estimated Creation Date" in the carbon dating service) as the article's carbon date. (See Figure 3)

Take in Figure 3

Hence we define "oldest variant", as follows:

> **Oldest variant** = the variant of paper found via Google Scholar that has the oldest carbon year date

We hence define "free citing window (oldest variant)" as follows.

> **Free citing window (oldest variant)** = Year in which citing paper was published - Year in which cited article (oldest variant) was made freely available

# Results

We took all available journal articles[2] listed on the Singapore Management University's Institutional Repository for the schools of Economics and School Sciences published in the year 2015.

We then obtained the full text of each paper and then sampled 2 citations each randomly from each paper. (See Table 2)

*Table 2: Number of citations sampled & percentage found free in 2016 via Google Scholar*

|  | No of journal articles | Total number of citations checked | Number of papers with at least one free variant in Google Scholar | % Free |
|---|---|---|---|---|
| School of Economics | 47 | 94 | 75 | 79.8% |
| School of Social Sciences | 24 | 48 | 25 | 50.0% |
| Total | 71 | 142 | 100 | 70.4% |

**Descriptive Statistics: Citation age for sampled cited papers (Economics and Social Sciences)**

*Table 3: Descriptive Statistics for citation age of sampled cited articles (Economics and Social Sciences)*

| Variable | N | StDev | Minimum | Median | Maximum |
|---|---|---|---|---|---|
| Citation Age | 142 | 11.104 | 0 | 8 | 56 |

**Frequency table: Citation age for sampled cited papers (Economics and Social Sciences)**

*Table 4: Frequency table: Citation age for sampled cited papers (Economics and Social Sciences)*

| Citation Age | N | % | Cum % |
|---|---|---|---|
| 0 | 5 | 3.52 | 3.52 |
| 1 | 7 | 4.93 | 8.45 |
| 2 | 18 | 12.68 | 21.13 |
| 3 | 7 | 4.93 | 26.06 |
| 4 | 13 | 9.15 | 35.21 |
| 5 | 6 | 4.23 | 39.44 |

---

[2] Out of 56 journal articles found under the School of Economics, only 47 journal articles had full text. Likewise, out of 95 journal articles found under the School of Social Sciences, only 24 journal articles had full text. We drop these articles without full text (and hence reference lists).

| 6 | 9 | 6.34 | 45.77 |
|---|---|------|-------|
| 7 | 3 | 2.11 | 47.89 |
| 8 | 9 | 6.34 | 54.23 |
| 9 | 2 | 1.41 | 55.63 |
| 10 | 7 | 4.93 | 60.56 |
| 11 | 1 | 0.70 | 61.27 |
| 12 | 7 | 4.93 | 66.20 |
| 13 | 6 | 4.23 | 70.42 |
| 14 | 6 | 4.23 | 74.65 |
| 15 | 4 | 2.82 | 77.46 |
| 16 | 3 | 2.11 | 79.58 |
| 17 | 2 | 1.41 | 80.99 |
| 18 | 2 | 1.41 | 82.39 |
| 19 | 3 | 2.11 | 84.51 |
| 20 | 2 | 1.41 | 85.92 |
| 22 | 2 | 1.41 | 87.32 |
| 23 | 1 | 0.70 | 88.03 |
| 24 | 3 | 2.11 | 90.14 |
| 26 | 3 | 2.11 | 92.25 |
| 27 | 1 | 0.70 | 92.96 |
| 28 | 1 | 0.70 | 93.66 |
| 29 | 1 | 0.70 | 94.37 |
| 30 | 1 | 0.70 | 95.07 |
| 32 | 1 | 0.70 | 95.77 |
| 34 | 1 | 0.70 | 96.48 |
| 45 | 1 | 0.70 | 97.18 |
| 47 | 1 | 0.70 | 97.89 |
| 55 | 1 | 0.70 | 98.59 |
| 56 | 2 | 1.41 | 100.00 |

Our samples do not include any articles published in Gold Journals. They also tend to be pretty old having a median citation age of 8 years. (See Table 3 & 4).  This seems to be the nature of the discipline, as studies have shown that median citation age in Economics can be as high as 10 years(Waller, 2006).

Of the citations we sampled, we ran a search in Google Scholar with their titles in the months of May and June 2016 to check if there were free copies found via Google Scholar. 79.8% (n=75) and 50.0%(n=25) of sampled cited papers (Economics and Social Science respectively) had at least one free copy found via Google Scholar in May-Jun 2016. (See Table 2).

As you can see from Table 5 and Table 6, most of the cited articles that were found to be free (Economics) were pretty old, with a median citation age of 8 years. A similar result can be seen for social sciences with a median citation age of 7 years for cited articles that were found to be free (table excluded for reasons of space).

While it can be argued that older articles are more likely to be free, as any embargo period is likely to have ended, we find in a subsample analysis of cited papers with citations age of 2 or less, 87.5%(n=16) and 100% (n=2) of them (Economics and Social Science respectively) had at least one free copy found via Google Scholar in May-Jun 2016. So the amount of free material cannot be simply explained by citation age.

**Descriptive Statistics: Citation age for cited papers (Economics) – Free versions available**

*Table 5: Descriptive Statistics: Citation age for cited papers (Economics) – Free versions available*

| Variable | N | StDev | Minimum | Median | Maximum |
|---|---|---|---|---|---|
| Citation Age | 75 | 10.45 | 0 | 8 | 56 |

**Frequency table: Citation age for cited papers (Economics) – Free versions available**

*Table 6: Frequency table for citation age of freely available cited articles (Economics)*

| Citation Age | Citation Year | N | % | Cum % |
|---|---|---|---|---|
| 0 | 2015 | 4 | 5.33 | 5.33 |
| 1 | 2014 | 4 | 5.33 | 10.67 |
| 2 | 2013 | 8 | 10.67 | 21.33 |
| 3 | 2012 | 4 | 5.33 | 26.67 |
| 4 | 2011 | 6 | 8.00 | 34.67 |
| 5 | 2010 | 2 | 2.67 | 37.33 |
| 6 | 2009 | 3 | 4.00 | 41.33 |
| 7 | 2008 | 2 | 2.67 | 44.00 |
| 8 | 2007 | 5 | 6.67 | 50.67 |
| 9 | 2006 | 2 | 2.67 | 53.33 |
| 10 | 2005 | 5 | 6.67 | 60.00 |
| 11 | 2004 | 1 | 1.33 | 61.33 |
| 12 | 2003 | 2 | 2.67 | 64.00 |
| 13 | 2002 | 5 | 6.67 | 70.67 |
| 14 | 2001 | 4 | 5.33 | 76.00 |
| 15 | 2000 | 2 | 2.67 | 78.67 |
| 16 | 1999 | 2 | 2.67 | 81.33 |
| 17 | 1998 | 1 | 1.33 | 82.67 |
| 18 | 1997 | 1 | 1.33 | 84.00 |
| 19 | 1996 | 3 | 4.00 | 88.00 |
| 22 | 1993 | 1 | 1.33 | 89.33 |
| 23 | 1992 | 1 | 1.33 | 90.67 |
| 24 | 1991 | 2 | 2.67 | 93.33 |

| 26 | 1989 | 2 | 2.67 | 96.00 |
| 27 | 1988 | 1 | 1.33 | 97.33 |
| 56 | 1959 | 2 | 2.67 | 100.00 |

For each free variant copy found in Google Scholar we record the URL and downloaded the article found. Each article URL was then carbon dated via http://cd.cs.odu.edu/ .

93.3% (n=70) and 76.0% (n=19) of sampled cited papers (Economics and Social Science respectively) had at least one variant that could be carbon dated.

Due to the small absolute number of carbon dated cited articles from social sciences, we drop further analysis on the samples for Social Sciences and focus on the samples from Economics.

Below shows the analysis for the free citing window (oldest variant) (See Table 7& 8)

### Descriptive Statistics: Free citing window (oldest variant) (Economics)

*Table 7: Descriptive statistics for free citing window (oldest variant) (Economics)*

| Variable | N | StDev | Minimum | Median | Maximum |
|---|---|---|---|---|---|
| Free citing window (oldest variant) | 70 | 6.857 | -1 | 7 | 17 |

### Frequency table: Free citing window (oldest variant) (Economics)

*Table 8: Frequency table for free citing window (oldest variant) (Economics)*

| Free citing window of Oldest Variant (In Years) | N | % | Cum % |
|---|---|---|---|
| >0 | 1[3] | 1.428571 | 1.428571 |
| 0 | - | - | - |
| 1 | 5 | 7.142857 | 8.571429 |
| 2 | 6 | 8.571429 | 17.14286 |
| 3 | 6 | 8.571429 | 25.71429 |
| 4 | 4 | 5.714286 | 31.42857 |
| 5 | 6 | 8.571429 | 40 |

---

[3] In our sample, we found one cited paper that was made free in 2016, hence there is a negative citing window here.

| | | | |
|---|---|---|---|
| 6 | 4 | 5.714286 | 45.71429 |
| 7 | 9 | 12.85714 | 58.57143 |
| 8 | 7 | 10 | 68.57143 |
| 9 | 4 | 5.714286 | 74.28571 |
| 10 | 1 | 1.428571 | 75.71429 |
| 11 | 5 | 7.142857 | 82.85714 |
| 12 | 7 | 10 | 92.85714 |
| 13 | 2 | 2.857143 | 95.71429 |
| 14 | 1 | 1.428571 | 97.14286 |
| 15 | 1 | 1.428571 | 98.57143 |
| 16 | - | - | - |
| 17 | 1 | 1.428571 | 100 |

We observed that many of the cited papers from Economics were made available for a long period before the citing paper was published. More than 50% have a free citing window of 7 years or more (See Table 8). This implies 50% of free cited articles were made free more than 8 years the cited papers were published (i.e. 2015).

## Discussion

A study of SHERPA/RoMEO publishers over 12 years have showed that restrictions on when, who and how self-archiving may take place has increased 119%, 190% and 1000% respectively. (Gadd and Troll Covey 2016). This does not seem to have affected our sample of cited papers probably because our sample consisted of old papers, such that even lengthy embargo periods would have expired.

While the long citation windows can be explained by authors in Economics tendency to cite old papers that were published long ago (Waller 2006), this doesn't tell the full story. Even if we restricted the analysis to only include cited papers with a citation age of 2 year or less, the average Free citing window (oldest variant) remains relatively high at 4 years meaning that even for the newer published papers, more than half are available 4 years before citing paper was published. This is a result of the widely held practice in Economics to circulate preprints and postprints months if not years before the paper is published.

Even after estimating when a paper was made free, we still face a question. While we can tell when an author publishes a paper, there is no easy way to determine when he actually began the task of doing his literature review, and may have needed access to the paper he eventually cited.

However given the high free citing windows (oldest variant) of most of the sampled cited papers it is highly likely the authors of the citing paper could have relied on these freely available paper if they wanted to.

For example, various studies have put publication lags in the field of Economics, Econometrics or Economics/Business at an average of 18.9 months (Yohe 1980), 22.8 months (Trivedi 1993) or over 16 months (Björk and Solomon 2013).

Assuming the most extreme case with a publication lag of 31.4 months found in the studies above (Trivedi 1993) and 12 months of research & writing, even this extreme case would take between 3-4 years to complete from writing to publication. Even in that extreme example, 74.3% of the cited papers that were found free today in 2016 could be used.

**Where the free papers were found**

As a secondary analysis, we also studied the domains of which the free papers were found.

*Table 9: Location types on which oldest free variants were found*

| Type | Count of Oldest Domains | % |
|---|---|---|
| Aggregator | 5 | 7.2% |
| Institutional Repository | 1 | 1.4% |
| Miscellaneous | 13 | 18.6% |
| Social Networking Websites for Researchers | 3 | 4.3% |
| Subject Repository | 1 | 1.4% |
| University Websites | 47 | 67.1% |
| Grand Total | 70 | 100.0% |

Table 9 shows an analysis of the types of organization hosting the oldest freely available item. We can see the majority of cited articles that were made freely available for the longest periods are from University websites (Table 9). Looking at the full URLs these seem to be pdfs put up by facility on their University web space for courses and other purposes. Unfortunately institutional repositories do not seem to factor at all into this.

It's possible the oldest papers were placed prior to the time where Institutional Repositories were made widely available or widely known. It's also interesting to notice that subject repositories like SSRN or RePEc do not seem to be represented in our sample.

*Table 10: Location types on which youngest free variants were found*

| Youngest Domains | Count of Youngest Domains | % |
|---|---|---|
| Aggregator | 4 | 5.7% |
| Government | 1 | 1.4% |
| Miscellaneous | 6 | 8.6% |
| Social Networking Websites for Researchers | 22 | 31.4% |
| Subject Repository | 3 | 4.3% |
| University Website | 34 | 48.6% |
| Grand Total | 70 | 100% |

We now turn to where the free papers which were put up most recently were found. (Table 10) While University Websites continue to be most common, we see Social Networking websites for researchers like ResearchGate start to play a part, perhaps reflecting the increasing popularity of such sites among researchers. This also supports the results found in more recent studies that show ResearchGate and its peers are becoming a popular place for researchers to deposit papers(Jamali and Nabavi, 2015, Scott and Sandra, 2014).

**What version of the paper is free?**

So far, we have been implicitly assuming that all free versions found are exactly the same. Of course any free version found via Google Scholar can be the pre-print, post-print or the published version (also known as version of record).

An attempt was made to distinguish between the types of free paper found. Due to the difficulty of distinguishing between pre-print and post-print versions of papers, they are both grouped together.

Amongst the oldest variants,

*Table 11: Type of free paper found (oldest variant)*

| Versions | N | % |
|---|---|---|
| Final | 23 | 32.85714 |
| NA | 2 | 2.857143 |
| Pre/Post | 45 | 64.28571 |
| Total | 70 | 100 |

We find that of the 70 oldest variants found, 64.3% (n=45) were either pre-print or post-print, with the remaining 32.8% (n=23) as the final published version. (See Table 11)

Would this affect the usability of the free version that a majority of the papers found were not final published version?

There isn't much research available on whether researchers are happy to access and cite pre-prints and post-prints. We know that in a survey of researchers done in 2008, respondents stated that when they had no access to the final version, 52.7% would never access the self-archived version (Morris and Thorn 2009). However the survey was done prior to the rise in popularity of Google Scholar and opinions expressed in survey may not reflect actual behavior.

Content wise though it is possible there might be little differences between the final published version and other versions. For example when comparing various similarity measures for the pre-prints in ArXiv vs published version there was very little differences in title, abstract or body full-text (Klein et al. 2016).

Overall, it's unclear if researchers would be happy enough with access to pre-prints and post-prints that can be found and forgo the final published version. If they insist on using the final published version then the fact that only 32% were final published versions would mean most of the freely available papers would not be used.

# Conclusion, limitations & Future work

We add to the growing literature on the availability of freely available articles online by taking into account when an article was made free allowing us to predict if a citing author could have used the freely available paper. By "carbon dating" freely available cited papers found on the Internet, we show that over 50% of such articles in the area of Economics were made available 7-8 years before the citing article was published. Even if we restricted our analysis to cited papers that were 2 years or newer, 50% of such articles were made available 3-4 years before the citing article was made available.

Given the typical publication lags of 2-3 years in Economics, we show that it is likely the authors of the citing papers could have used the freely available articles.

Given that faculty currently place great importance on the role of academic libraries in the purchaser role of books and journal articles overs other roles such as information literacy or discovery (Housewright et al., 2013) , the increased citation of freely available articles with long free citation windows can serve as a rough proxy on how important this role is and can have future implications on the future of academic libraries.

We also find that of the oldest articles made available, 67.1% of them were found on University sites (excluding Institutional repositories), showing that Institutional repositories are not commonly used at least in the past for the oldest freely available items. We find that of the newest papers just put up recently, ResearchGate is starting to rise in popularity.

Of freely available articles found, 32.8% were final published version. Further analysis could be done to study if such articles were made available legally and to see if such papers are associated with certain sites.

One of the limitations of this study is that it focuses mainly on the use of Google Scholar to surface free articles. With the recent rise in interest in Scihub(Elbakyan and Bohannon, 2016) faculty reliance of the library for access to articles may further weaken. Another limitation is that the study focuses mainly on the discipline in Economics, which has a tradition of citing older articles and long periods of releasing preprints and post-prints before publication and this combines to produce a high percentage of full text with long free citation windows. Other disciplines which focus on citing newer articles and/or may not have a tradition of preprints and postprints might produce shorter free citation windows.